\documentclass[journal]{IEEEtran}
\ifCLASSINFOpdf
\else
\usepackage[dvips]{graphicx}
\usepackage{algorithmic}
\usepackage{algorithm}
\graphicspath{{../eps/}}
\fi
\usepackage{amsmath}
\ifCLASSOPTIONcompsoc
  \usepackage[caption=false,font=normalsize,labelfont=sf,textfont=sf]{subfig}
\else
  \usepackage[caption=false,font=footnotesize]{subfig}
\fi
\begin{document}
%

\title{Efficient algorithm based on non-backtracking matrix for community detection in signed networks}

%
%
%


\author{
\IEEEauthorblockN{Zhaoyue~Zhong\IEEEauthorrefmark{2},  Xiangrong~Wang\IEEEauthorrefmark{3}\IEEEauthorrefmark{4}, Cunquan~Qu\IEEEauthorrefmark{1}\IEEEauthorrefmark{2}\thanks{Corresponding author: cqqu@sdu.edu.cn (C.Qu)}, Guanghui~Wang\IEEEauthorrefmark{1}\IEEEauthorrefmark{2}}
  
\IEEEauthorblockA{\IEEEauthorrefmark{1}Data Science Institute, Shandong Univeristy, China} 

\IEEEauthorblockA{\IEEEauthorrefmark{2}School of Mathematics, Shandong University, China} 

\IEEEauthorblockA{\IEEEauthorrefmark{3}Institute of Future Networks, Southern University of Science and Technology, Shenzhen, China} 

\IEEEauthorblockA{\IEEEauthorrefmark{4}Research Center of Networks and Communications, Peng Cheng Laboratory, Shenzhen, China} 
}

%
%

\markboth{Journal of \LaTeX\ Class Files,~Vol.~14, No.~8, August~2015}%
{Zhong \MakeLowercase{\textit{et al.}}: Bare Demo of IEEEtran.cls for IEEE Journals}
%



\maketitle

\begin{abstract}
Community detection or clustering is a crucial task for understanding the structure of complex systems. In some networks, nodes are permitted to be linked by either "positive" or "negative" edges; such networks are called signed networks. Discovering communities in signed networks is more challenging than that in unsigned networks. 
In this study, we innovatively develop a non-backtracking matrix of signed networks, theoretically derive a detectability threshold for this matrix, and demonstrate the feasibility of using the matrix for community detection. We further improve the developed matrix by considering the balanced paths in the network (referred to as a balanced non-backtracking matrix). Simulation results demonstrate that the algorithm based on the balanced non-backtracking matrix significantly outperforms those based on the adjacency matrix and the signed non-backtracking matrix. The proposed (improved) matrix shows great potential for detecting communities with or without overlap. 
\end{abstract}

\begin{IEEEkeywords}
Community detection, signed networks, non-backtracking matrix, spectral analysis, detectability threshold.
\end{IEEEkeywords}

%
\IEEEpeerreviewmaketitle

\section{Introduction}

\IEEEPARstart{C}{ommunities}, also known as clusters or modules, are groups of nodes that may share common attributes or have similar properties in a graph. Community detection involves division of similar nodes or nodes with many (positive, large weighted) connections into a group, thereby providing a possible approach for controlling the network. Since nodes with many (positive) connected edges often have similar properties, in terms of graphs, community detection is also a process of finding cut edges. If a few edges are removed, the network can be divided into several parts, and then, the division of these parts is, to a certain extend, equivalent to community partition. 

Community detection is widely applied in fields such as biology, computer science, engineering, economics, political science, and sociology \cite{fortunato2010community}. For example, protein-protein interaction networks are a research hotspot in biology and bioinformatics \cite{jonsson2006cluster}. The interaction between proteins is the basis of every process in the cell. Each interaction is observed experimentally and marked as a connection. Proteins with the same or similar functions are grouped into one module and are expected to participate in the same process. This is a classical application that conceptualizes the actual scenario as an unsigned network. A social network is also a typical example of a network with a community structure. In general research, a connection in as social network is regarded as being positive, e.g., followers, likes, and message forwarding. However, social networks often contain many negative connections. For example, some websites such as  \emph{epinions.com} and  \emph{slashdot.com} permit users to identify friends and enemies \cite{anchuri2012communities}. The signed network introduced in the present study represents such a scenario, i.e., opposite opinions on the same topic \cite{anchuri2012communities} and  blackout and reporting among users. 

\begin{figure}[ht]
	\centering
	\includegraphics[width = 0.9\linewidth]{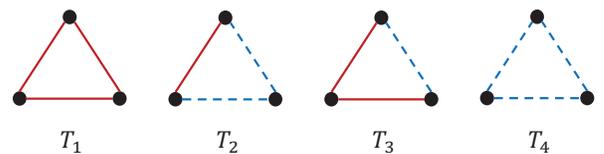}
	\caption{\textbf{Triangles in signed networks.} $T_1$ and $T_2$ are balanced and relatively stable. $T_3$ and $T_4$ are unbalanced and hence likely to break apart. }
	\label{fig:triangle}
\end{figure}

In the 1940s \cite{heider1946attitudes}, Heider introduced the concept of signed networks and proposed the well-known structural balance theory, which states "the friends of my friends, as well as the enemies of my enemies, are my friends"(see in Fig. \ref{fig:triangle}). This theory, which is one of the most popular social science theories has recently been receiving increasing attention. A related research topic is to design algorithms for computing the structural balance of large-scale datasets \cite{facchetti2011computing,marvel2011continuous,kirkley2019balance}. Another research topic is the study of the impact of structural balance on some concrete applications, such as recommender systems \cite{monika2018structural} and dynamic processes \cite{qu2019impact}.

In social networks, user communities provide valuable services for websites, such as friend recommendations for users. Clustering of web pages can be used to rank them and provide more relevant search results \cite{fortunato2010community}. Furthermore, the use of community detection in social media can clearly explain the observed phenomena and provide benchmarks for social mechanisms \cite{leskovec2010signed}.

The following are some of the categories into which existing method for community detection can typically be classified: (1) conventional algorithms such as graph partitioning \cite{kernighan1970efficient}, hierarchical clustering \cite{friedman2001elements}, partition clustering \cite{macqueen1967some}, and spectral clustering \cite{von2007tutorial}; (2) modular-based methods \cite{yang2016modularity}; (3) dynamic algorithms \cite{girvan2002community,newman2004finding}; and (4) methods based on statistical inference \cite{mackay2003information}. 
Most of these algorithms are for unsigned networks. However, community detection in signed networks is a more challenging task, because of the distinct roles of the negative inter-community and intra-community links. In general, negative inter-community links segregate the connected communities whereas negative the inter-community links blur the community structure.

In this study, we mainly consider the spectrum method for detecting the community structure. For community detection in unsigned networks, the adjacency matrix \cite{chauhan2009spectral}, Laplacian matrix \cite{pothen1997graph}, non-backtracking matrix \cite{krzakala2013spectral}, and other structure-related matrices have been used, whereas for community detection in signed networks, the adjacency matrix has been explored \cite{morrison2019community}.

The remainder of this paper is organized as follows. First, in Sec.\ref{sec: non-backtracking}, we define the necessary notations and propose the definition of the non-backtracking matrix. We take full advantage of the structural balance theory to propose the signed non-backtracking (SNBT) matrix of signed networks. In Sec.\ref{sec:community detection}, we derive a theoretical detection threshold $ \mu_{c}>\sqrt{c}$ (where $ \mu_{c} $ is a community-correlated eigenvalue and $c$ is the average degree of the network) and theoretically demonstrate the feasibility of using the SNBT matrix for community detection above the detectability threshold. In Sec.\ref{sec:algorithm}, we introduce an improved version of the SNBT matrix, termed the balanced non-backtracking (BNBT) matrix, and propose an efficient community detection algorithm based on the BNBT matrix. In Sec.\ref{sec: results}, we present numerical simulations performed for demonstrating the effectiveness of the SNBT matrix vis-{\`a}-vis the adjacency matrix in community detection. Through experiments on signed stochastic block networks and real-world networks, we show that BNBT matrix-based approach has the best performance among the three approaches based on these three matrices. Finally, in Sec.\ref{sec: conclusions}, we conclude the paper.

\section{Non-backtracking matrix for signed networks}\label{sec: non-backtracking}

\subsection{Signed stochastic block model}

Signed networks consist of interacting individuals with both positive and negative relationships. Each individual in the network corresponds to a node in the graph. The connection between a pair of individuals is regarded as the edge between the corresponding node pair. Positive and negative relationships are represented as positive and negative edges in the network. For simplicity, the weights of positive and negative edges are defined as $ 1 $ and $ -1 $, respectively.

First, we describe the signed stochastic block model \cite{krzakala2013spectral,morrison2019community,decelle2011asymptotic}. Given an undirected network with $ n $ nodes (where $ n $ is assumed to be an even number), we divide the node set into two groups, $ \mathcal{A} $ and $ \mathcal{B} $, where each group has $ n/2 $ nodes. Nodes in group $ \mathcal{A} $ are indexed from $ 1 $ to $ n/2 $  and those in group $ \mathcal{B} $ are indexed from $ n/2+1 $ to $ n $.

A signed network can be represented by an adjacency matrix $ A $ in which the entries take on values of $ \left\{1,-1 , 0\right\} $, where 0 signifies the absence of an edge, and $1$ and $-1$ signify a positive relationship and negative relationship, respectively. The adjacency matrix is symmetric because the network is undirected. The following are some of the parameters for any pair of nodes $ (i,j) $. The probability of formation of an edge between any given in-group (out-group) node pair is $ d_{in}  $ ($ d_{out} $). The expected edge density of the entire network is $ d=d_{in}+d_{out} $. Given the presence of an edge between in-group members, $p_{in}^+$ denotes the conditional probability that it will be positive. Analogously, $ p_{out}^{+} $, $ p_{in}^{-} $, and $ p_{out}^{-} $ denote the conditional probabilities of a positive edge between out-group members, a negative edge between in-group members, and a negative edge between out-group members, respectively. Thus, the conditional probabilities satisfy $ p_{in}^{+}+p_{in}^{-}=1 $ and $ p_{out}^{+}+p_{out}^{-}=1 $. 

Generally speaking, when we say that a network has a community structure, at least one of the following conditions holds:
\begin{enumerate}
    \item \label{case:1}  The number of positive intra-community links is greater than the number of negative links, i.e., $p_{in}^+>p_{out}^+$ and $p_{in}^{-}<p_{out}^{-}$.
    \item \label{case:2}  The density of intra-community links is higher than that of inter-community relationships, i.e., $d_{in} > d_{out}$. 
\end{enumerate}

Under the first condition, the community structure is termed relationship-sensitive and is denoted as \emph{r-community}. Under the second condition, the community structure is termed link-density-sensitive and  denoted as \emph{d-community}.  

\subsection{Definition of non-backtracking matrix}

One of the main contributions of this work is to define non-backtracking matrices for signed networks, which have great potential for community detection. Though a non-backtracking matrix of unsigned networks is well defined--which is presented herein for completeness--a proper definition of a non-backtracking matrix is far from trivial, as demonstrated in following sections. 

Prior to providing a formal definition of a non-backtracking matrix of signed networks, we first present the definition of a non-backtracking matrix of unsigned  or general networks. 
The non-backtracking matrix $\widetilde{H}$, often termed the $Hashimoto$ matrix in mathematics, is defined as follows: 
\begin{equation}\label{def:hashimoto}
    \widetilde{H}_{(u\rightarrow v),(w\rightarrow x)}= \widetilde{A}_{(u,v)}\widetilde{A}_{(w,x)}\mathbf{1}(v=w)\mathbf{1}(u\neq x),
\end{equation}
where $\widetilde{A}$ is the adjacency matrix of the unsigned network and $e=(u,v)$ and $f=(w,x)$ are two directed edges. It should be noted that if the network is undirected, we treat each undirected edge as two directed edges. Hence, the matrix $\widetilde{H}$ has the dimensions $2m\times 2m$, where $m$ is the number of edges in the network.

The non-backtracking matrix $ \widetilde{H}_{2m\times2m} $ can, in fact, be expressed as 
\begin{equation}
 \widetilde{H}_{(u\rightarrow v),(w\rightarrow x)} =\begin{cases}
1 &\text{if}\ v=w\ \text{and}\ u\not=x,\\
0 &\text{otherwise}.
\end{cases}
\end{equation}

Similar to the matrix $\widetilde{H}$ defined in Eq.(\ref{def:hashimoto}), the SNBT matrix of signed networks, denoted as $ H $, can be directly derived as follows:
\begin{equation}\label{def:signhashimoto}
H_{(u\rightarrow v),(w\rightarrow x)}= A_{(u,v)}A_{(w,x)}\mathbf{1}(v=w)\mathbf{1}(u\neq x).
\end{equation}

This matrix can be expressed as
\begin{equation}
H_{(u\rightarrow v),(w\rightarrow x)} = 
\begin{cases}
1 &\text{if} \ v=w, u\not=x \ \text{and}\\ &\sigma(u\rightarrow v)=\sigma(w\rightarrow x),
\\
-1 &\text{if} \ v=w, u\not=x \ \text{and}\\ &\sigma(u\rightarrow v)\not=\sigma(w\rightarrow x),
\\
0 &\text{otherwise},
\end{cases} 
\end{equation}
where $ \sigma\left(u\rightarrow v \right)$ denotes the sign of a directed edge $ u \rightarrow v $, which takes a value of either $ 1 $ or $ -1 $.
The significance of the defined non-backtracking matrix $ H $ is that true information will be transferred between two edge pairs with identical signs and false information will be transferred between two edge pairs with different signs, and this behavior accurately follows the structural balance theory (a triple with either one or three negative signs is unstable). 

On the basis of the information flowing along the directed edges in the network, $H$ can also be deduced via linearized belief propagation (LBP). This is explained in greater detail in Appendix \ref{subsec:BP}.

In other words, we define the non-backtracking matrix from two different perspectives, theoretically deduce its role in community detection, and confirm the feasibility of its application to a basic stochastic block model via the linearization of the updating equation of the BP algorithm around a fixed point.

\section{Theoretical analysis} \label{sec:community detection}

\subsection{Analytical derivation of community detection threshold and detection vector}

To demonstrate the applicability of the SBNT matrix to community detection, we derive the community detection threshold and a detection vector for an arbitrary signed network.
By generalization on the basis of observations in unsigned networks \cite{krzakala2013spectral,ren2010graph,angel2015non,kotani20002}, we define $ g^{out} $ and $ g^{in} $ as $ n $-dimensional vectors:
\begin{equation}
g^{out}_{u}=\sum_{v\in \mathcal{N}(u)}g_{u\rightarrow v}\cdot \sigma(u\rightarrow v),\nonumber
\end{equation}
\begin{equation}
g^{in}_{u}=\sum_{v\in \mathcal{N}(u)}g_{v\rightarrow u}\cdot \sigma(v\rightarrow u),\nonumber
\end{equation}
where $ \mathcal{N}(u) $ denotes the neighbor set of node $ u $ and vector $g$ is a given vector with 2m dimensions. Unlike in the case of  unsigned network, we not only sum over incoming and outgoing edges but also consider the sign of the edges.

By applying $ H $ to $ g $, we get 
\begin{equation}
(Hg)_{u}^{out}=\sum_{v\in \mathcal{N}(u)}g_{v}^{out}-g_{u}^{in},\nonumber
\end{equation}
\begin{equation}
(Hg)_{u}^{in}=(k_{u}-1)\sum_{v\in \mathcal{N}(u)}g_{v}^{out},\nonumber
\end{equation}
where $ k_{u} $ denotes the degree of node $ u $ (which is independent of the sign of the edges).
Rewriting the above two equations in matrix form gives
\begin{equation}
\begin{pmatrix}
(Hg)^{in}\\
(Hg)^{out}
\end{pmatrix}
=H' 
\begin{pmatrix}
g^{in}\\
g^{out}
\end{pmatrix},\nonumber
\end{equation}
\begin{equation}
H'=\begin{pmatrix}
0 & D-I\\
-I & \widetilde{A}
\end{pmatrix},
\end{equation}
where $ I $ is the identity matrix, $ D $ is the diagonal matrix of vertex degrees, and $ \widetilde{A} $ is the adjacency matrix of the underlying unsigned structure corresponding to the signed network.

Suppose that $ Hg=\mu g $; then, we have
\begin{equation}
\mu\begin{pmatrix}
g^{in}\\
g^{out}
\end{pmatrix}=H'\cdot \begin{pmatrix}
g^{in}\\
g^{out}
\end{pmatrix}.\nonumber
\end{equation}
If $ g^{in} $  and $ g^{out} $  are nonzero, then $ \begin{pmatrix}
g^{in}\\
g^{out}
\end{pmatrix} $  is an eigenvector of $ H' $ with the same eigenvalue $ \mu $. Hence,
\begin{equation}
\mu g^{out}=\widetilde{A}\cdot g^{out}-g^{in}=[\widetilde{A}-\mu ^{-1}(D-I)]g^{out}.\nonumber
\end{equation}
Therefore, $ \mu $ is a root of the quadratic eigenvalue equation
\begin{equation}\label{eq:zeta}
det\left| \mu ^{2}I-\mu \widetilde{A}+(D-I) \right| =0.
\end{equation}

The complexity of calculating the eigenvalues of  $ H' $  will be much lower than that of the original non-backtracking matrix $H$. Eq.(\ref{eq:zeta}) is well known in the theory of graph zeta functions \cite{kotani20002}. It accounts for $ 2n $ of the eigenvalues of $ H $, and the remaining $ 2m-2n $ eigenvalues are $ \pm 1 $.

In fact, we directly and simply prove that the spectrum of $ H $ is the same as that of $ \widetilde{H} $, where the network for the latter matrix is considered as an unsigned network. $ H $ can also be derived via multiplication of all the elements of some rows and their corresponding columns in $ \widetilde{H} $ by $ -1 $, that is,
\begin{equation}
det\left|\lambda I-H\right|=(-1)^{2\left|\epsilon_{-}\right|}\cdot det\left|\lambda I-\widetilde{H}\right|=det\left|\lambda I-\widetilde{H}\right|,
\end{equation}
where $ \left|\epsilon_{-}\right| $ is the number of negative edges in the network.

Therefore, the bulk of the spectrum $ B $ is also confined to the radius disk $ \sqrt{c} $ in signed networks. It should be noted that  $ c=d\cdot n $ is the average degree of the network. We can similarly and correspondingly define $ c_{in} $ and $ c_{out} $.

Further, we obtain the first and second eigenvalues of $ H $ as
\begin{equation}
\begin{cases}
\mu_{1}\approx c\\
\mu_{2}\approx\mu_{c}=\dfrac{c_{in}-c_{out}}{2}=\dfrac{d_{in}-d_{out}}{2}n
\end{cases}.
\end{equation}

In the unsigned network, the second eigenvector of the non-backtracking matrix is a community-correlated eigenvector. If the second eigenvalue of $ \widetilde{H} $ is separated from the bulk of the spectrum, then the eigenvector corresponding to the second eigenvalue can be used in the community detection (via labeling of the vertices according to the sign of the sum of all incoming edges at each vertex) \cite{janson2004robust}. Similar conclusions for signed networks are verified subsequently in the paper.

\begin{figure}[ht]
	\centering
	\includegraphics[width=0.6\linewidth]{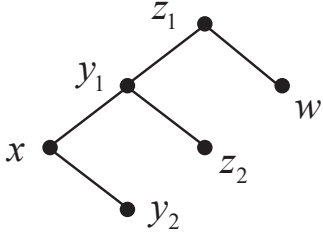}
	\caption{\textbf{Illustration of calculation of $ d(u\rightarrow v,w\rightarrow x) $.} A transverse of two edges $y_{1}\rightarrow z_{1}  $ and $ z_{1}\rightarrow w $ needs to be considered to go from edge $ x \rightarrow y_1 $ to edge $ z_{1}\rightarrow w $. Thus, $ d(x\rightarrow y_{1},z_{1}\rightarrow w)=2 $.}
	\label{fig:cal_distance}
\end{figure}

Next, we first attempt to construct a vector $ g $ that is correlated to the communities and is an approximate eigenvector with eigenvalue $ \mu_{c} $. We assume that $ c=O(1) $, and therefore the graph is sparse and locally tree-like. For any positive integer $ r $ and any directed edge $ (u,v) $, we define the following:
\begin{equation}
g_{u\rightarrow v}^{(r)}=\mu_{c}^{-r}\cdot \sum_{(w,x):d(u\rightarrow v,w\rightarrow x)=r}\sigma_{x}\cdot \sigma_{u\rightarrow v},\nonumber
\end{equation}
where $ \sigma_{x}=\pm 1 $ denotes the community of $x$, $ \sigma_{u\rightarrow v}=\pm 1 $ denotes the sign of edge $ (u,v) $, and $ d(u\rightarrow v,w\rightarrow x) $  denotes the number of steps required to go from $ u\rightarrow v $ to $ w\rightarrow x $ in the graph of directed edges, as shown in Fig. \ref{fig:cal_distance}.

Application of $ H $ to $ g^{(r)} $ gives
\begin{equation}
(Hg^{(r)})_{u\rightarrow v}=\mu_{c}^{-r}\cdot \sum_{(w,x):d(u\rightarrow v,w\rightarrow x)=r+1} \sigma_{x}\cdot \sigma_{u\rightarrow v},\nonumber
\end{equation}
which can be simplified as
\begin{equation}\label{key}
(Hg^{(r)})_{u\rightarrow v}=\mu_{c}\cdot g_{u\rightarrow v}^{(r+1)}\nonumber
\end{equation}
We can write $ g_{u\rightarrow v}^{(r)}-g_{u\rightarrow v}^{(r+1)} $ as
\begin{equation}
\mu_{c}^{-r}\cdot \sigma_{u\rightarrow v}\cdot \sum_{(w,x):d(u\rightarrow v,w\rightarrow x)=r}(\sigma_{x}-\mu_{c}^{-1}\sum_{y\in N(x)\setminus w}\sigma_{y}).\nonumber
\end{equation}
Now, there are (an expected) $ c^{r} $ terms in this sum, each of which is conditioned on $ \sigma_{x} $ values and has an expected value of zero and a constant variance. Hence,
\begin{equation}
E[(g_{u\rightarrow v}^{(r)}-g_{u\rightarrow v}^{(r+1)})^{2}]=O(c^{r}\mu_{c}^{-2r}).\nonumber
\end{equation}

By summing over all the edges, we obtain
\begin{equation}
E[(g^{(r)}-g^{(r+1)})^{2}]=O(c^{r}\mu_{c}^{-2r}\left|E\right|).\nonumber
\end{equation}
Therefore, when the community-correlated eigenvalue (the second eigenvalue) satisfies
\begin{equation}\label{fomula20}
\mu_{c}>\sqrt{c},
\end{equation}
it can be naturally considered that when this eigenvalue is separated from the bulk spectrum, the error is small and it approaches zero for large $ r $.
Further, from the conclusion for unsigned networks \cite{mossel2003information,kesten1966additional},  it can be inferred that, under the condition that the threshold is met and $ n\rightarrow \infty $, for every $ u\rightarrow v $, 
\begin{equation}
<g_{u\rightarrow v}^{(r)},\sigma_{u}\cdot \sigma_{u\rightarrow v}>\ne 0.\nonumber
\end{equation}
Thus, we can draw the following conclusion:
\begin{equation}
\left|Hg^{(r)}-\mu_{c}g^{(r)}\right|=o(1).\nonumber
\end{equation}

Therefore, $ g^{(r)} $ is indeed an approximate eigenvector of $ H $ having an eigenvalue of $ \mu_{c} $, which may be used to detect the community structure of signed networks. Further, Eq.(\ref{fomula20}) expresses the detection threshold finally derived in this study, which shows agreement with the threshold for unsigned networks.

\subsection{Threshold for the case of more than two communities}\label{subsec: beyond2}

The above-presented analysis is for stochastic block models with two communities ($ q=2 $). In fact, according to the above-described derivation process, the SNBT matrix can also be well applied to a model with more than two communities ($ q>2 $). Its detectable threshold should be similar to the conclusion for the unsigned network; that is, the community-correlated eigenvalue satisfies Eq.(\ref{fomula20}), i.e., $ \mu_{c}>\sqrt{c} $.

In this case, the second eigenvalue is given as
\begin{equation}
\mu_{2}\approx\dfrac{c_{in}-c_{out}}{q}=\dfrac{d_{in}-d_{out}}{q}n.\nonumber
\end{equation}
We can take the first $ k $ eigenvectors and use the \slshape k-means \upshape algorithm to determine the group labels of the nodes.
However, the threshold value for detection of the network structure using the adjacency matrix when the number of communities is greater than two is yet unknown; therefore, we refrain from performing a more detailed comparative evaluation here.

\section{Algorithm for community detection}\label{sec:algorithm}
\subsection{Improved non-backtracking matrix for community detection in signed networks}

From the above analysis, we know that the SNBT matrix is density sensitive rather than sign sensitive. Even in the case where the negative edges of inter-community connections and the positive edges of intra-community connections constitute the majority of the edges in the connections, the final results are not ideal as long as the density of connected edges within and between groups exceeds the above-derived threshold value. This insensitivity of the SNBT matrix to edge signs is essentially contrary to our original goal of exploring the community structure of signed networks

In fact, from the below-described threshold of the adjacency matrix, we can see that this matrix is only sign-sensitive and that is, too, does not meet our requirements for community detection. The ideal matrix should have good performance in terms of both aspects; that is, it should be both sign-sensitive and density-sensitive. 

We considered both the structural balance theory and the BP theory in the construction of the non-backtracking matrix. Why does this sign insensitivity occur? We find from the discussion in Sec.\ref{sec:community detection}A that the approximation of the second eigenvector is actually related only to the $k$-order neighbors of node $x$, which is simply equal to the total number of communities the node belongs to and whether or not the path between two nodes is balanced, i.e., whether the internal relationship is friendly or hostile, is not considered.

Hence, we improve the matrix by making the following assumption. We assume that a large number of balanced or stable paths of a given length $k$ between two vertices $u$ and $v$ is expected if both the vertices belong to the same community. Then, the improved matrix, denoted by $H^b$, is defined as follows: 
\begin{equation}
    H^b_{(u\rightarrow v),(w\rightarrow x)} = \mathbf{1}(v=w)\mathbf{1}(u\neq x)\mathbf{1}(A_e\cdot A_f=1).
\end{equation}
This matrix can alternatively be expressed as 
\begin{equation}\label{eq:balanced_non}
H^b_{(u\rightarrow v),(w\rightarrow x)} = 
\begin{cases}
1 &\text{if} \ v=w, u\not=x \ \text{and}\\
&\sigma(u\rightarrow v)=\sigma(w\rightarrow x),
\\
0 &\text{otherwise}.
\end{cases} 
\end{equation}

Therefore, according to the BP algorithm, during the propagation process, only true information is transmitted to edges with same signs, and false information is transmitted to edges with different signs. For simplicity, we term this matrix the BNBT matrix. In fact, $H^b$ is an approximation of $H$, and the detection threshold for community detection is still unclear. However, the below-described experiments demonstrate that this improved matrix is ideal for community detection.

\subsection{Community detection algorithms}

In this section, we provide a detailed description of the community detection algorithm based on the BNBT matrix. We calculate the first $q-1$ eigenvectors of the adjacency matrix for performing clustering using the $k$-means algorithm.
Because the detection accuracy of BNBT matrix is not validated theoretically and the community-related eigenvectors of the BNBT matrix are also unclear, we perform clustering using all eigenvectors corresponding to the real eigenvalues. When applying the SNBT-matrix-based algorithm, we calculate the first $\min\{q,r\}$, where $r$ is the number of real eigenvalues.

\begin{algorithm}[htb] 
\caption{ Framework for community detection in signed networks.} 
\label{alg:Framwork} 
\begin{algorithmic}[1] 
\REQUIRE ~~\\ 
Adjacency matrix of network $A$;\\
Number of communities $q$;\\
\ENSURE ~~\\ 
\STATE The BNBT matrix $H^b$ is constructed according to Eq.(\ref{eq:balanced_non});
\label{code:construct_bnon }
\STATE The eigenvectors  $[\xi_2,\xi_3,\dots,\xi_r]$ corresponding to the real eigenvalues, where $r$ is the number of real eigenvalues, are calculated; 
here, each $(r-1)$th-row vector represents a directed link;
\label{code:edge_eigenvectors}
\STATE For each node $u$, weighted summation of all the vectors of its out-edges $(u,v)$, $v\in \mathcal{N}(u)$ is performed to obtain its representative vector;
\label{code:node_eigenvector}
\STATE Then, clustering is performed using the $k$-means algorithm to obtain the group label $\widetilde{g}$ of the nodes; 
\label{code:clustering}
\RETURN $\widetilde{g}$. 
\end{algorithmic}
\end{algorithm}

In the following section, we compare the community detection performances of the three matrices.

\section{Results}\label{sec: results}

To evaluate the community detection performances of the SBNT matrix ($H$) and BNBT matrix ($H^b$), we perform extensive simulations on signed networks. The accuracy of community detection is quantified by the concept of overlap \cite{krzakala2013spectral}, which is defined as the ratio of the number of correctly predicted nodes to the total number of nodes. Overlap can be expressed as
\begin{equation}
ovl = \dfrac{1}{n}\sum_{u}\delta_{g_{u},\widetilde{g}_{u}},\nonumber
\end{equation}
where $ g_{u} $ is the actual group label of vertex $ u $, and $ \widetilde{g}_{u} $  is the label inferred by the algorithm. When $ g_{u}=\widetilde{g}_{u} $ for every node $ u $, we have $ovl=1$ and the detection accuracy is 100$\%$.

We break symmetry by maximizing the overall $ q! $ permutations of the groups, where $q$ is the number of groups into which the nodes are divided. The prediction is accurate when the overlap equals $ 1 $, and under this definition, the minimum value of the overlap can be taken as $ 1/q $. The overlap is normalized as
\begin{equation}
ovl = (\dfrac{1}{n}\sum_{u}\delta_{g_{u},\widetilde{g}_{u}}-\dfrac{1}{q})/(1-\dfrac{1}{q}).\nonumber
\end{equation}

The overlap ranges from 0 to 1, where a value of  $0$ implies that the prediction is inaccurate because of random grouping. For visualization purposes, we still use the un-normalized overlap in the below-described numerical simulations.

\subsection{Comparison with adjacency-matrix-based detection}\label{section4.2}
In unsigned networks, most of the eigenvalues of the adjacency matrix are within a threshold. The number of eigenvalues beyond the threshold equals the number of communities. The second eigenvector indicates the community structure\cite{krzakala2013spectral}.
\begin{figure}[ht]
	\centering
	\includegraphics[width=0.9\linewidth]{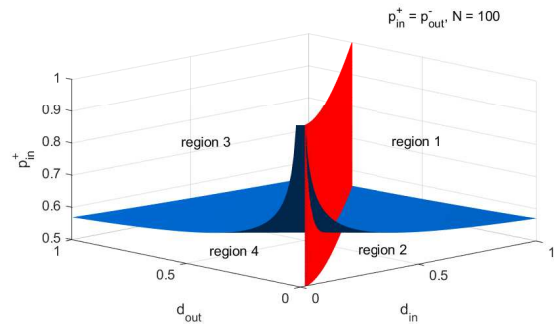}
	\caption{\textbf{Detection threshold of non-backtracking matrix.} $ p_{in}^{+}=p_{out}^{-} $ and $n=100$. The red surface indicates the boundary that can be detected by the proposed (points in regions 1$\&$2 can be detected), whereas the blue surface indicates the boundary that can be detected by the adjacency-matrix-based algorithm (points in regions 1$\&$3 can be detected).}
	\label{fig:thrsh_nonvsadj}
\end{figure}
Similar results have been reported in \cite{morrison2019community}, that is, the bulk of the spectrum of the adjacency matrix is also within a threshold for signed networks. Unlike in the case of the unsigned network, in the signed network, the leading eigenvector represents the community structure, and the number of eigenvalues beyond the threshold is no longer equal to the number of communities (it should be equal to the number of communities minus 1).
Moreover, the authors of \cite{morrison2019community} were the first to consider the detectability transition in signed networks. They concluded that when there are only two communities, as long as the following conditions (Eq.66 in \cite{morrison2019community}) are met, the sign of the principal eigenvector can be used to detect communities using perturbation analysis and random matrix theory.

\begin{equation}\label{eq:thrsh_adj}
\begin{split}
p_{out}^{-}>&\dfrac{1}{2}-\dfrac{d_{in}}{d_{out}}\left(p_{in}^{+}-\dfrac{1}{2}\right)\\
&+\dfrac{1}{d_{out}}\sqrt{\dfrac{d_{in}+d_{out}-8d_{in}^{2}\left(p_{in}^{+}-\dfrac{1}{2}\right)^{2}}{2n}}
\end{split}.
\end{equation}

Considering the average degree $c$ of the signed network, we use the SNBT matrix as long as the following inequality is satisfied:
\begin{equation}\label{eq:thrsh_NBT}
\begin{cases}
\dfrac{1}{2}<p_{in}^{+}<\dfrac{1}{2}+\dfrac{1}{2}\sqrt{\dfrac{c}{c^{2}+nd_{in}^{2}}}\\
d_{in}>\dfrac{c+\sqrt{c}}{n}
\end{cases}.
\end{equation}

\begin{figure*}[h]
	\centering 
	\includegraphics[width = 0.9\linewidth]{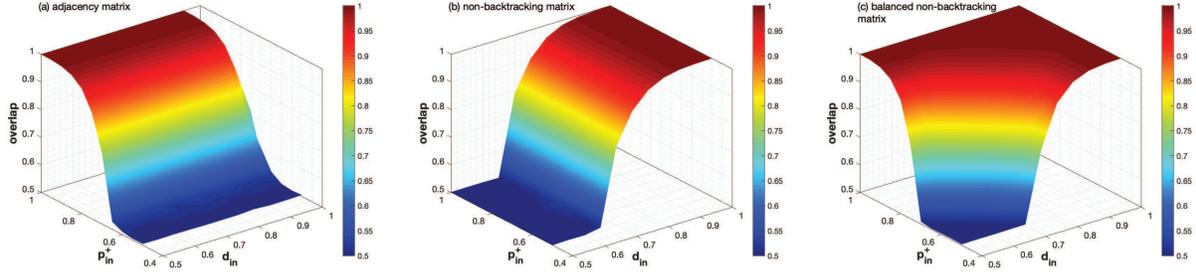} 
	\caption{\textbf{Community detection performances of three different matrices.} $n = 10000$, $c=10$, $d_{in} + d_{out}=1$, and $p_{in}^+ = p_{out}^-$. $d_{in}$ and $ p_{in}^+$ both vary from $0.5$ to $1$ with an interval of $0.05$.      }
	\label{fig:compare_matrices}
\end{figure*}

In some cases, we can obtain better results by applying the SNBT matrix than by applying the adjacency matrix. Here, we provide a simple but general example for comparing the two methods (see Fig. \ref{fig:thrsh_nonvsadj}). In all the comparisons, we assume $ p_{in}^{+}=p_{out}^{-} $ for convenience. We only consider the case of $ p_{in}^{+}>0.5 $ here, because when $ p_{in}^{+}<0.5 $, the community structure can be represented by $ u_N $ (the last eigenvector of the adjacency matrix). However, given the dynamic evolution process of the network, only the driving role of the leading eigenvector of the initial network in the evolution of structural balance yields a dynamic manifestation of the detectability transition \cite{morrison2019community}.

In fact, Fig. \ref{fig:thrsh_nonvsadj} is an abstract representation of the community related parameters in a real network. Region 4 represents the situation wherein $ d_{in}<d_{out}$, $p_{in}^{+}\approx p_{in}^{-}$ and $p_{out}^{+}\approx p_{out}^{-}$. In other words, the random blocks generated by the parameters in region 4 are not the two communities mentioned above. For the case that the parameters lie in region 2 (or region 3), we actually anticipate that the algorithm is not sign-sensitive (or density-sensitive), and in such a scenario, the BNBT matrix is effective.
 
It should be noted that because the theoretical detection threshold of the BNBT matrix is still unknown, we refrain from discussing it any further. 

\subsection{Comprehensive comparison among three matrices in terms of community detection performance}
\begin{figure*}[h]
	\centering 
	\includegraphics[width = 0.6\linewidth]{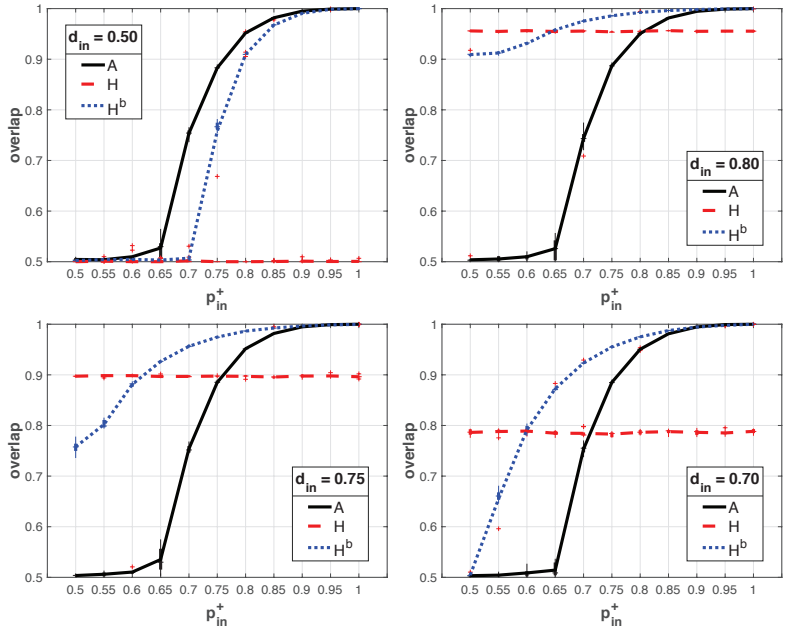} 
	\caption{\textbf{Comparison of three matrices with fixed $d_{in}$.} $n=10000$, $c=10$, and $d_{in}^+ = 0.50, 0.80, 0.75$, and  $0.70$. $p_{in}^+$ varies from $0.5$ to $1$ with an interval of $0.05$.     }
	\label{fig:compare_varpin}
\end{figure*}

In the following, we provide examples to demonstrate the superior performance of the improved algorithm. The comparison results are shown in Fig. \ref{fig:compare_matrices}. The $ x $-axis and $y$-axis represents the $ p_{in}^{+} $ and  $d_{in}$, respectively. Both the $z$-axis and the color represent the overlap. The experiments are performed on signed networks with a size of $10^4$ and an average degree of $10$. We set $p_{in}^{+}=p_{out}^{-}$ for convenience, which varies from $0.5$ to $1$ with an interval of $0.05$. It is worth noting that the $d_{in}$ values in the figure are normalized, i.e., $d_{in}+d_{out}=1$, which also varies from $0.5$ to $1$ with an interval of $0.05$. 
\begin{figure*}[ht]
	\centering 
	\includegraphics[width = 0.8\linewidth]{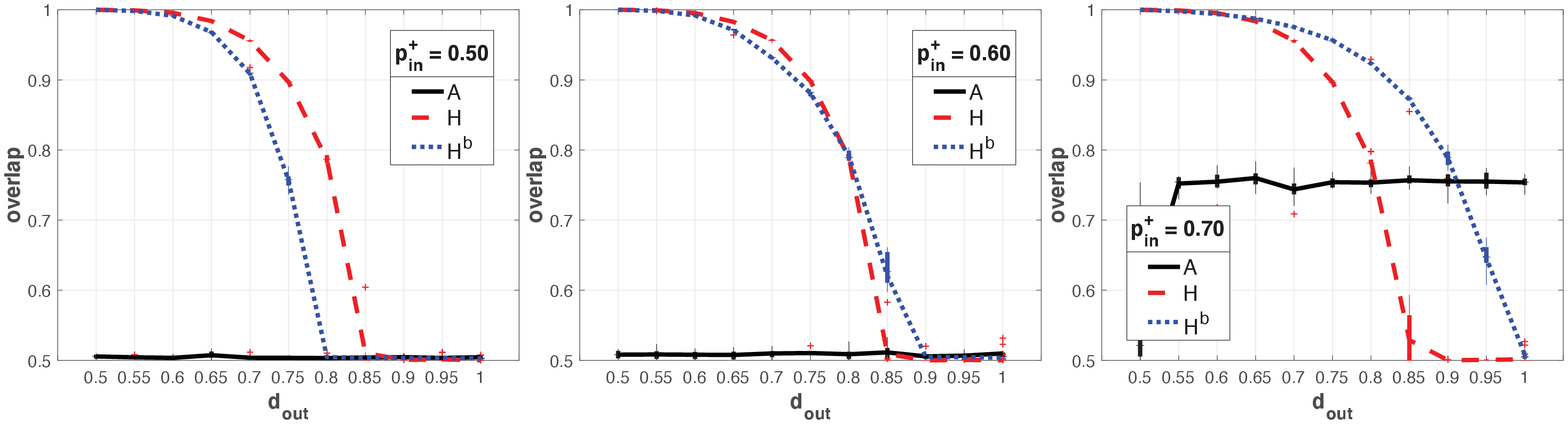} 
	\caption{\textbf{Comparison of three matrices with fixed $p_{in}^+$.} $n=10000$, $c=10$, and $p_{in}^+ = 0.50, 0.60$, and $0.70$. $d_out$ varies from $0.5$ to $1$ with an interval of $0.05$.      }
	\label{fig:compare_vardin}
\end{figure*}
We can see that the performance of the SNBT matrix is sensitive only to the link density of inter- and intra- community connections, whereas the adjacency-matrix-based algorithm is more sensitive to the link signs. The BNBT-matrix-based algorithm has the advantages of both these matrices, and its performance depends on both the link density and the link signs. Moreover, the area of the undetectable field (in which the overlap is about $0.5$) is smallest, which indicates that this matrix has the best performance.

Fig. \ref{fig:compare_varpin} shows four concrete examples with $d_{in} \in \{0.5, 0.70, 0.75, 0.80\}$ and varying $p^{+}_{in}$ (in the range of [$0.5$, $1$]). In fact, each line in this figure is a slice of the data in shown Fig. \ref{fig:compare_matrices}, which is obtained by considering the corresponding $d_{in}$ values. The black, red, and blue lines represent the results obtained using the adjacency matrix, the SNBT matrix, and the BNBT matrix, respectively. For each case (i.e., for each matrix), we perform 10 experiments and calculated the average value for joining the data points to form the black, blue, and red dashed lines. From all the graphs except for the first one, which is the case of $d_{in}=0.5$, we can make the following observations. The overlap is close to 0.5 (which means that the nodes are labelled almost randomly) when the detection threshold of the adjacency matrix is not met, however, the algorithms based on the two non-backtracking matrices show good performances. As discussed above, the overlap does not change with an increase in $p_{in}^+$. As $p_{in}^+$ increases, the adjacency matrix-based algorithm outperforms the SNBT-matrix-based algorithm. However, the BNBT-matrix-based algorithm outperforms the adjacency-matrix-based approach at all $p_{in}^+$ values. This entire analysis can lead to a satisfactory conclusion regarding the superiority of the BNBT-matrix-based algorithm. The first graph is the only exception; in this case, the density is of no use for dividing the community structure($d_{in}=d_{out}=0.5$), and therefore, the SNBT-matrix-based algorithm is invalid and the performance of the BNBT-matrix-based algorithm is poorer than that of the sign-sensitive adjacency-matrix-based algorithm.

As shown in Fig. \ref{fig:compare_vardin}, we further evaluate the performances of the three approaches from another perspective; that is, we provide three concrete examples with $p_{in}^+ \in \{0.50, 0.60, 0.70\}$ and varying $d_{out}$ (in the range of $[0.5,1]$). These data lines are slices obtained from the data in Fig. \ref{fig:compare_matrices} from a different angle. The performance of adjacency-matrix-based approach is not completely independent of $d_{out}$. When $p_{in}^+=0.7$, the overlap increases at a small $d_{out}$. However, after the threshold is met, the performance of this approach does not improve with decreasing $d_{out}$. It is further proved that the SNBT matrix $H$ is sign-insensitive. As was the case in the above-described analysis, density sensitivity is the only factor we need to consider when $p_{in}^+=0.5$, and therefore, $H$ is superior; however, at other $p_{in}^+$ values, $H^b$ is superior.

In addition, we know that detection of community structure may be difficult when the graph is sparse. For example, in an unsigned network, when $ c $ is constant and $ n $ is large, the network is decomposed for several reasons. Most importantly, the leading eigenvalues of $ A $ are represented by maximum-degree vertices, and the corresponding eigenvectors are localized around these vertices \cite{krzakala2013spectral,krivelevich2003largest}. The non-backtracking matrix has better performance in the case of a sparse graph. We expect it to have superior performance even in the case of a signed network.

If the right side of the first inequality in Eq.(\ref{eq:thrsh_NBT}) is regarded as a function of $ d_{in} $, we obtain a lower bound according to the monotonicity of the function as follows:
\begin{equation}
\begin{cases}
\dfrac{1}{2}<p_{in}^{+}<\dfrac{1}{2}+\dfrac{1}{2}\sqrt{\dfrac{1}{2c}}\\
d_{in}>\dfrac{c+\sqrt{c}}{N}
\end{cases}.
\end{equation}

\begin{figure}[ht]
	\centering
	\includegraphics[width=0.6\linewidth]{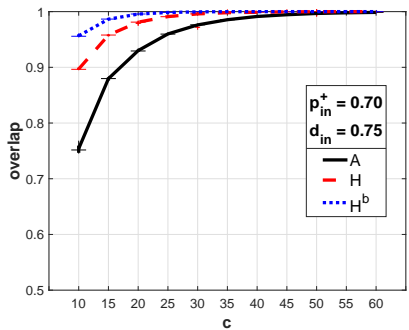}
	\caption{\textbf{Performances of three matrices for graphs with different levels of sparsity.} $n=10^4\text{, } p^+_{in}=0.7\text{, and } d_{in}=0.75$ .}
	\label{fig:overlap_c}
\end{figure}

Therefore, we theoretically prove that when the community detection using the non-backtracking matrix is feasible, the smaller the $c$ value, the better are the results obtained using the non-backtracking matrices than those obtained the adjacency matrix. It should be noted that what we refer to as better performance in the case of a sparse graph is relative. In fact, with an increase in the sparsity of the network, the detection accuracy will indeed decrease correspondingly.
The observation is also confirmed by numerical simulations performed on stochastic block models with $n=10^4$, $p^+_{in}=0.7$, $d_{in}=0.75$, and varying average degree $c$ (in the range of $[10,60]$ with an interval of $5$). As shown in Fig. \ref{fig:overlap_c}, the performances of the two non-backtracking matrix-based algorithms under varying $c$ are more stable and robust than that of the adjacency-matrix-based algorithm.

We are also interested in the performance of the newly proposed $H^b$ in the above comparative analysis; however, these analyses are based only on numerical simulations, and the underlying theory remains to be studied. Nevertheless, we can draw a definitive conclusion that $H^b$ is sensitive to both the edge sign and the connection density. In most cases, it has better performance than the other two matrices, $A$ and $H$. These two matrices have their own limitations. Although they may be the best choice in some very special scenarios, generally speaking, the BNBT-matrix-based algorithm is the most reliable and effective approach.

\subsection{Experiments on real-world networks}

\begin{table}[h]
  \caption{Basic features of three real-world networks with ground-truth communities.}
  \label{tab:basic_feature}
\begin{center}
    \begin{tabular}{cccc}
      \hline
      \hline
      Network & $n$ & $m$ & $q$  \\
      \hline
      Email-Eu-core\cite{leskovec2007graph,yin2017local} & 1,005 & 25,571 & 42 \\
      \hline
      American College football\cite{girvan2002community} & 115 & 613 & 12 \\
      \hline
      Political blogs\cite{adamic2005political} & 1,490 & 19,090 & 2\\
      \hline
      \hline
    \end{tabular}
  \end{center}
\end{table}

\begin{figure*}[h]
	\centering 
	\includegraphics[width =0.8\linewidth]{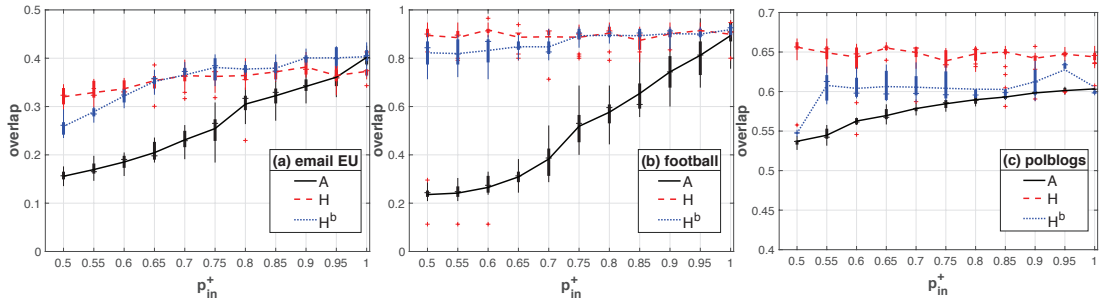} 
	\caption{\textbf{Comparison of community detection performance of three matrices for real-world networks} $p_in^{+}$ varies from $0.5$ to $1$ with an interval of $0.05$.       }
	\label{fig:real_world}
\end{figure*}

Finally, we select three real-world networks with ground-truth communities to compare the community detection performances of the three matrices. The basic features of the three networks are listed in Table \ref{tab:basic_feature}. However, since these networks are unsigned, we use the following method to assign a sign (positive or negative) to each link. First, we classify all the links into two categories: inter-community links and intra-community links. Then, we assign to each inter-community link a positive sign with a probability of $p^+_{out}$ and to each intra-community link a probability of $p_{in}^+$. We set $p^+_{out}+p^+_{in} = 1$, which means that the larger the $p^+_{in}$ value, the more significant is the community structure of the network. 

In Fig. \ref{fig:real_world}, we have illustrated the community detection accuracies of the three matrices under variation of $p^+_{in}$ from $0.5$ to $1$ with an interval of $0.05$. As can be seen from this figure, the performance of the adjacency-matrix-based algorithm increases with increasing $p_{in}^+$. This result reveals that this algorithm is sign-sensitive, which is in agreement with the earlier discussion. Similarly, the accuracy of the SNBT-matrix-based algorithm remains almost unchanged, because it is insensitive to $p_{in}^+$. The BNBT-matrix-based algorithm shows the best performance and its accuracy also increases with increasing $p^+_{in}$. In addition, even though the number of communities in the football network is $12$, the BNBT- and SNBT-matrix-based methods still perform well. This result implies that these matrices are applicable for community detection in a signed network when the number of communities is greater than two, which further confirms our observations.

\section{Conclusion and discussion}\label{sec: conclusions}

In this study, we investigate efficient community detection in a signed network by demonstrating the feasibility of defining a non-backtracking matrix of signed networks. We provide the definition of an appropriate non-backtracking matrix from two different perspectives: the well known structural balance theory and belief propagation. We analytically determine the community detection ability of the proposed non-backtracking matrix and propose an even more efficient matrix, termed the balanced non-backtracking matrix $H^b$; the algorithm based on this matrix significantly outperforms that based on the adjacency matrix. 

We anticipate the algorithm based on the proposed balance non-backtracking matrix to be sensitive and adaptable to both the edge sign and the connection density. Our algorithm (as also the previous algorithms) is completely effective in the case of the most standard community partition; however, we need to have a basic understanding of more complex networks and communities with different characteristics before we can select the most appropriate algorithm for them. From this viewpoint, the balanced non-backtracking matrix is the most universal matrix. An exception to this universality is that when the community in the actual network is not a relationship-dependent community or a density-dependent community, the above-discussed algorithms may not provide satisfactory results.

The proposed framework shows great potential for community detection with or without overlap and paves the way to understanding the collective behaviors of systems in which positive and negative relationships coexist.


%

\section*{Acknowledgment}
Z-YZ, C-QQ, and G-HW were supported in part by National Natural Science Foundation of China under Grant 12001324, Grant 11631014, and Grant 11871311, in part by China Postdoctoral Science Foundation under Grant 2019TQ0188, Grand 2019M662315, in part by Shandong University multidisciplinary research and innovation team of young scholars under Grand 2020QNQT017. X-RW acknowledges the partial support of the project "PCL Future Greater-Bay Area Network Facilities for Large-scale Experiments and Applications (LZC0019)".

\ifCLASSOPTIONcaptionsoff
  \newpage
\fi



%
\bibliographystyle{IEEEtran}
\bibliography{cas-refs.bib}

\appendices

\section{Alternative definition of non-backtracking matrix based on linearized belief propagation}\label{subsec:BP}

Belief propagation (BP) is a kind of acyclic message passing algorithm, which calculates the exact marginal distribution of each vertex in the network. Although BP is designed to work accurately on trees, it is usually applied to general graphs that are sparse and  may contain loops \cite{krzakala2013spectral,ren2010graph,coja2009spectral}.

This algorithm starts from the appropriate initial assignment and performs an iteration for some "messages". Specifically, for each edge $ (v,w) $ in a graph $ G= (V,E) $, the message $ \eta_{v\rightarrow w}^{a} $ indicates the conditional probability that $ v $ belongs to community $ a $ when $ w $ does not, and the message $ \eta_{w\rightarrow v}^{a} $ indicates the probability that $ w $ belongs to community $ a $ when $ v $ does not. Usually, $ \eta_{v\rightarrow w}^{a}\ne\eta_{w\rightarrow v}^{a} $. As is clear from this explanation, although the original graph is undirected, these messages are delivered to the directed edge, where each message has a value between 0 and 1. The message can be calculated iteratively on the basis of such information transfer.

The BP algorithm has good consistency with the actual group allocation because it approximates the Bayesian optimal reasoning of a block model. The application of the BP algorithm to spectral clustering is also a possible research direction \cite{angel2015non,mellor2019graph}.

In this section, we prove that in signed networks, $ H $ appears in the linearization equation derived from the updating equation of the BP algorithm.
Because of the presence of the edge sign, we generalize the existing BP updating equation into the following form for $ u \in \mathcal{N}(v) $:
\begin{small}
\begin{equation}
\begin{split}
&\dfrac{\eta_{v\rightarrow w}^{+}}{\eta_{v\rightarrow w}^{-}}:=e^{-h}\\
&\times \dfrac{\prod_{\sigma(u\rightarrow v)=\sigma(v\rightarrow w)}\left(\eta_{u\rightarrow v}^{+}c_{in}+\eta_{u\rightarrow v}^{-}c_{out}\right)}{\prod_{\sigma(u\rightarrow v)= \sigma(v\rightarrow w)}\left(\eta_{u\rightarrow v}^{+}c_{out}+\eta_{u\rightarrow v}^{-}c_{in}\right)}\\
&\times \dfrac{\prod_{\sigma(u\rightarrow v)\ne \sigma(v\rightarrow w)}\left(\left(1-\eta_{u\rightarrow v}^{+}\right)c_{in}+\left(1-\eta_{u\rightarrow v}^{-}\right)c_{out}\right)}{\prod_{\sigma(u\rightarrow v)\ne \sigma(v\rightarrow w)}\left(\left(1-\eta_{u\rightarrow v}^{+}\right)c_{out}+\left(1-\eta_{u\rightarrow v}^{-}\right)c_{in}\right)},\label{updating}
\end{split}
\end{equation}
\end{small}
where $ \eta_{v\rightarrow w}^{\pm} $ denotes the probability that $ v $ belongs to a community when $ u $ does not belong to the network and $ \pm $ denotes two separate communities. It should be noted that $ e^{-h} $  represents the information transmitted from non-edges (points not adjacent to v), where $ h=(c_{in}-c_{out} )(n_{+}^{BP}-n_{-}^{BP}) $, and $ n_{\pm}^{BP} $  denotes to the ratio of the current number of points in two communities to the total number of nodes estimated by the BP algorithm.

It is noteworthy that when $ u\rightarrow v $ and $ v\rightarrow w $ have different signs, the information transmitted to $ v\rightarrow w $ is not  $ \eta_{u\rightarrow v}^{\pm} $ but $ \left(1- \eta_{u\rightarrow v}^{\pm} \right) $. This means that if these two edges have different signs, the false information will be transmitted to $v\rightarrow w$.
Similarly, the trivial fixed point of the above updated equation is still $  \eta_{v\rightarrow w}=1/2 $; that is, the probability of each vertex being divided into two communities is equal.

Next, we consider the information update equation near the trivial fixed point.
By taking $ \eta_{u\rightarrow v}^{\pm}=1/2 \pm \delta_{u\rightarrow v}  $ and linearizing around this fixed point (for more details, see Appendix \ref{appB}), we obtain an updating rule of $ \delta $:
\begin{equation}
\delta:=\dfrac{\left(c_{in}-c_{out}\right)}{\left(c_{in}+c_{out}\right)}H^{T}\delta.
\end{equation}
That is, $H $ can also be obtained by the BP algorithm.

\begin{small}
\section{Derivation process of updating equation of $\delta$}\label{appB}

To simplify the updating equation of $\delta$, let us carefully consider Eq.$\left(\ref{updating}\right)$.

First, we notice that $ n_{+}^{BP}=n_{-}^{BP} $ holds near the trivial fixed point; that is, $ e^{-h} $ can be written as $1$. Under the assumption of an unknown constant $Z$, we split Eq.$\left(\ref{updating}\right)$ into two equations:
\begin{equation}
\begin{split}
\eta_{v\rightarrow w}^{+}=Z\cdot\prod_{\substack{u\in N(v)\\ \sigma(u\rightarrow v)=\sigma(v\rightarrow w)}}\left(\eta_{u\rightarrow v}^{+}c_{in}+\eta_{u\rightarrow v}^{-}c_{out}\right)\\
\times \prod_{\substack{u\in N(v)\\ \sigma(u\rightarrow v)\ne \sigma(v\rightarrow w)}}\left[\left(1-\eta_{u\rightarrow v}^{+}\right)c_{in}+\left(1-\eta_{u\rightarrow v}^{-}\right)c_{out}\right],\label{eq1}
\end{split}
\end{equation}

\begin{equation}
\begin{split}
\eta_{v\rightarrow w}^{-}=Z\cdot\prod_{\substack{u\in N(v)\\ \sigma(u\rightarrow v)=\sigma(v\rightarrow w)}}\left(\eta_{u\rightarrow v}^{+}c_{out}+\eta_{u\rightarrow v}^{-}c_{in}\right)\\
\times \prod_{\substack{u\in N(v)\\ \sigma(u\rightarrow v)\ne \sigma(v\rightarrow w)}}\left[\left(1-\eta_{u\rightarrow v}^{+}\right)c_{out}+\left(1-\eta_{u\rightarrow v}^{-}\right)c_{in}\right].\label{eq2}
\end{split}
\end{equation}

Rewriting Eq.$\left(\ref{eq1}\right)$ near the trivial fixed point $ \eta_{u\rightarrow v}^{\pm}=1/2 \pm \delta_{u\rightarrow v}  $ gives
\begin{small}
\begin{equation}
\begin{split}
&\frac{1}{2}+\delta_{u\rightarrow v}=Z\\
&\times \prod_{\substack{u\in N(v)\\ \sigma(u\rightarrow v)=\sigma(v\rightarrow w)}}\left[\left(\frac{1}{2}+\delta_{u\rightarrow v}\right)c_{in}+\left(\frac{1}{2}-\delta_{u\rightarrow v}\right)c_{out}\right]\\
&\times \prod_{\substack{u\in N(v)\\ \sigma(u\rightarrow v)\ne \sigma(v\rightarrow w)}}\left[\left(\frac{1}{2}-\delta_{u\rightarrow v}\right)c_{in}+\left(\frac{1}{2}+\delta_{u\rightarrow v}\right)c_{out}\right].
\end{split}
\end{equation}
\end{small}

By merging similar items, we get
\begin{equation}
\begin{split}
&\frac{1}{2}+\delta_{u\rightarrow v}=Z\\
&\times \prod_{\substack{u\in N(v)\\ \sigma(u\rightarrow v)=\sigma(v\rightarrow w)}}\left[\frac{1}{2}\left(c_{in}+c_{out}\right)+\delta_{u\rightarrow v}\left(c_{in}-c_{out}\right)\right]\\
&\times \prod_{\substack{u\in N(v)\\ \sigma(u\rightarrow v)\ne \sigma(v\rightarrow w)}}\left[\frac{1}{2}\left(c_{in}+c_{out}\right)-\delta_{u\rightarrow v}\left(c_{in}-c_{out}\right)\right].\label{eq3}
\end{split}
\end{equation}

Eq.$\left(\ref{eq2}\right)$ can be simplified to 
\begin{equation}
\begin{split}
&\frac{1}{2}-\delta_{u\rightarrow v}=Z\\
&\times \prod_{\substack{u\in N(v)\\ \sigma(u\rightarrow v)=\sigma(v\rightarrow w)}}\left[\frac{1}{2}\left(c_{in}+c_{out}\right)-\delta_{u\rightarrow v}\left(c_{in}-c_{out}\right)\right]\\
&\times \prod_{\substack{u\in N(v)\\ \sigma(u\rightarrow v)\ne \sigma(v\rightarrow w)}}\left[\frac{1}{2}\left(c_{in}+c_{out}\right)+\delta_{u\rightarrow v}\left(c_{in}-c_{out}\right)\right].\label{eq4}
\end{split}
\end{equation}

Linearization of Eq.$\left(\ref{eq3}\right)$ and Eq.$\left(\ref{eq4}\right)$ gives
\begin{small}
\begin{equation}
\begin{split}
&\frac{1}{2}+\delta_{v\rightarrow w} \approx Z\cdot\Bigg\{\left[\frac{1}{2}\left(c_{in}+c_{out}\right)\right]^{\left|N(v)\right|}\\
&+\sum_{\substack{u\in N(v)\\ \sigma(u\rightarrow v)=\sigma(v\rightarrow w)}} \delta_{u\rightarrow v}\left(c_{in}-c_{out}\right)\left[\frac{1}{2}\left(c_{in}+c_{out}\right)\right]^{\left|N(v)-1\right|}\\
&-\sum_{\substack{u\in N(v)\\ \sigma(u\rightarrow v)\ne \sigma(v\rightarrow w)}} \delta_{u\rightarrow v}\left(c_{in}-c_{out}\right)\left[\frac{1}{2}\left(c_{in}+c_{out}\right)\right]^{\left|N(v)-1\right|} \Bigg\},\label{eq5}
\end{split}
\end{equation}
\end{small}

\begin{small}
\begin{equation}
\begin{split}
&\frac{1}{2}-\delta_{v\rightarrow w} \approx Z\cdot\Bigg\{\left[\frac{1}{2}\left(c_{in}+c_{out}\right)\right]^{\left|N(v)\right|}\\
&-\sum_{\substack{u\in N(v)\\ \sigma(u\rightarrow v)=\sigma(v\rightarrow w)}} \delta_{u\rightarrow v}\left(c_{in}-c_{out}\right)\left[\frac{1}{2}\left(c_{in}+c_{out}\right)\right]^{\left|N(v)-1\right|}\\
&+\sum_{\substack{u\in N(v)\\ \sigma(u\rightarrow v)\ne \sigma(v\rightarrow w)}} \delta_{u\rightarrow v}\left(c_{in}-c_{out}\right)\left[\frac{1}{2}\left(c_{in}+c_{out}\right)\right]^{\left|N(v)-1\right|} \Bigg\}.\label{eq6}
\end{split}
\end{equation}
\end{small}

To eliminate the constant $Z$, we calculate the sum and difference of Eq.$\left(\ref{eq5}\right)$ and Eq.$\left(\ref{eq6}\right)$,
\begin{small}
\begin{equation}
\begin{split}
&1=2Z \left[\frac{1}{2}\left(c_{in}+c_{out}\right)\right]^{\left|N(v)\right|},\\
&2\delta_{v\rightarrow w}=2Z  \left[\frac{1}{2}\left(c_{in}+c_{out}\right)\right]^{\left|N(v)-1\right|}\times \left(c_{in}-c_{out}\right)\\
&\times \left(\sum_{\substack{u\in N(v)\\ \sigma(u\rightarrow v)=\sigma(v\rightarrow w)}} -\sum_{\substack{u\in N(v)\\ \sigma(u\rightarrow v)\ne \sigma(v\rightarrow w)}} \right)\delta_{u\rightarrow v} .
\end{split}
\end{equation}
\end{small}

After eliminating the constant $Z$, we have
\begin{equation}
\begin{split}
&\delta_{v\rightarrow w}=\frac{c_{in}-c_{out}}{c_{in}+c_{out}}\\
&\times \left(\sum_{\substack{u\in N(v)\\ \sigma(u\rightarrow v)=\sigma(v\rightarrow w)}}-\sum_{\substack{u\in N(v)\\ \sigma(u\rightarrow v)\ne \sigma(v\rightarrow w)}}\right)\delta_{u\rightarrow v}.
\end{split}
\end{equation}

Consequently, we obtain the updating rule of $ \delta $ in signed networks,
\begin{equation}
\delta:=\dfrac{\left(c_{in}-c_{out}\right)}{\left(c_{in}+c_{out}\right)}H^{T}\delta.
\end{equation}


\end{small}




%

\begin{IEEEbiography}[{\includegraphics[width=1in,height=1.25in,clip,keepaspectratio]{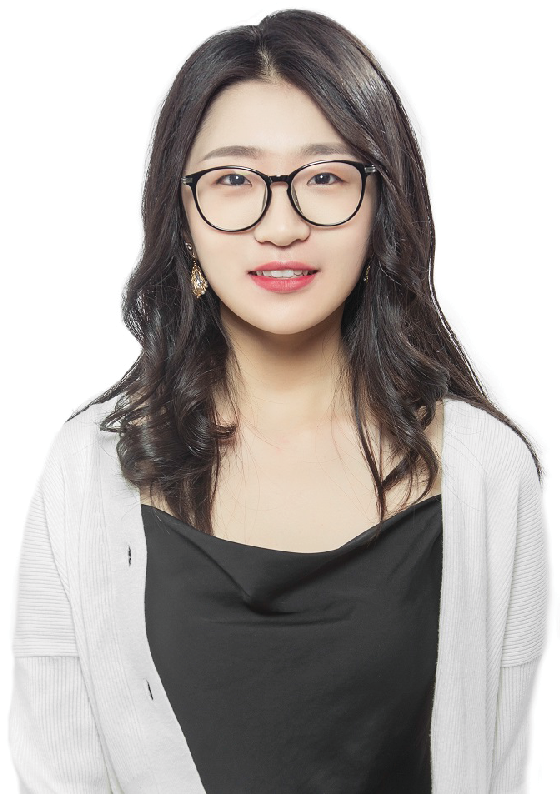}}]{Zhaoyue Zhong}received the B.S.  degree from School of Mathematics, Shandong University, China in 2020. She is currently pursuing the M.S. degree in School of Mathematical Sciences, Fudan University, China.
Her current research interests include complex networks and mathematical methods in neural networks.
\end{IEEEbiography}

\begin{IEEEbiography}[{\includegraphics[width=1in,height=1.25in,clip,keepaspectratio]{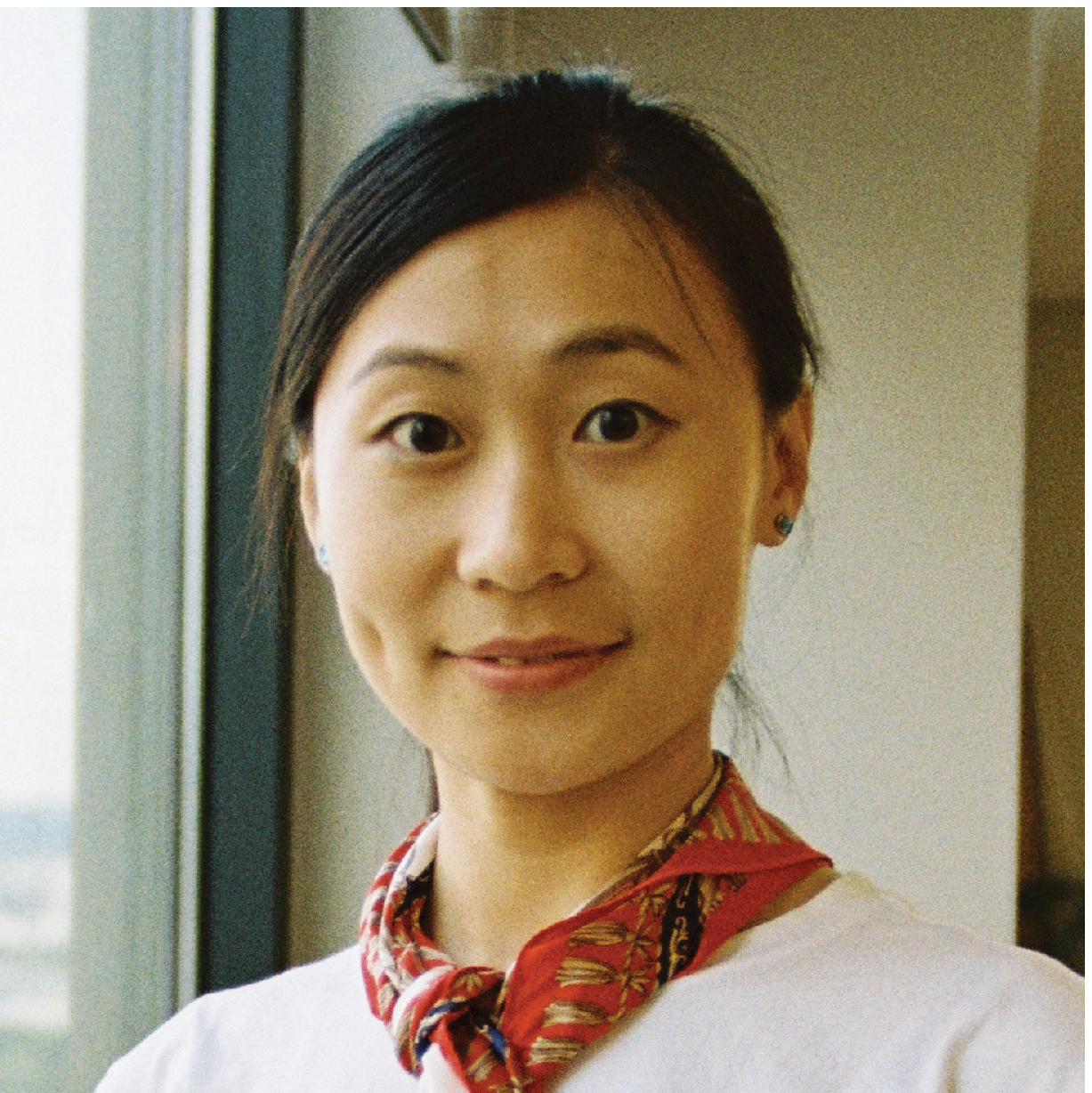}}]{Xiangrong Wang}
is currently a research assistant professor at Southern University of Science and Technology, Shenzhen, 
China. She received her Ph.D. degree from the Delft University of Technology, the Netherlands in 2016. Before joining Shenzhen, she was a postdoctoral researcher at the Delft University of Technology 
from 2017 to 2018. In 2017 and 2018, she was a visiting scholar at Zaragoza University, Zaragoza, Spain and ISI Foundation, Torino, Italy. Her research focuses on modeling and analysis of complex networks, nonlinear dynamics and graph spectral analysis.
\end{IEEEbiography}

\begin{IEEEbiography}[{\includegraphics[width=1in,height=1.25in,clip,keepaspectratio]{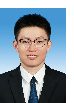}}]{Cunquan Qu}
is currently working as a Post-doctor researcher at Data Science Institute, Shandong University, Jinan, China. He did a joint Ph.D. project in the Multimedia Computing Group at the Delft University of Technology for two years. He received his Ph.D. degree from Shandong University in 2019. His research focuses on analyzing network structure, modeling the dynamics process, graph neural networks.
\end{IEEEbiography}

\begin{IEEEbiography}[{\includegraphics[width=1in,height=1.25in,clip,keepaspectratio]{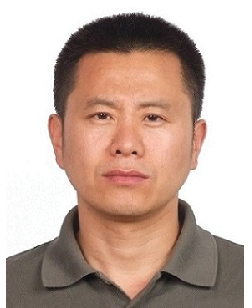}}]{Guanghui Wang}
received a B.Sc. degree in Mathematics from Shandong University, Jinan, China, in 2001. He got the doctor’s degree from Paris SDU University and worked in Ecole Centrale Paris as a Post-doctor. Now he is a professor in School of Mathematics, Shandong University. His current interests include graph theory, combinatorics, complex networks and bioinformatics.
\end{IEEEbiography}





\end{document}